\documentstyle{article}
\begin{document}

\author{Iftah Gideoni\thanks{%
Email: iftah@physics.ubc.ca}}
\title{Toward general solutions to time-series problems: Notes on obstacles and
noise}
\date{15 September 1995}
\maketitle

\begin{abstract}
Computational difficulties in the general application of Bretthorsts formalism to 
time-series problems, posed by the large number 
of possible models and the use of models with
nonorthogonal base-functions are discussed. 
The specific problem under consideration is a Bayesian procedure 
for model selection,  parameter estimation, and classification, that was 
applied to the search for the {\it {In Vivo}} $T_2$ decay rate 
distributions in brain tissues. Through the estimation of the 
meta-parameter $\sigma$ in the  process, we also gain a better understanding of 
the meaning and estimation of "noise" in the frame-work of probability 
theory as logic.
\end{abstract}

\section{Introduction}

Probability theory as logic establish the unique procedure with which one can 
find the probability assignment to any well-posed inference problem. 
Nonetheless, this procedure may still be difficult to compute.  In {\cite{BretBook}}, 
Bretthorst present a mathematical frame-work in which time-series problems can be 
treated: In this frame-work,
Each of our models is a linear combination of base functions. that is, the
model $M_\alpha $  as a function of the time $t,$ is of the form:

\begin{equation}
\label{modgenmod}M_\alpha (t)=\sum_{j=1}^{m_\alpha}B_{j\alpha}G_{j\alpha}(t,\overrightarrow{\tau
_{j\alpha}}) 
\end{equation}

Where:

\begin{itemize}
\item  $\overrightarrow{\tau _{j\alpha}}$ is the set of the nonlinear parameters
of the base function $G_{j\alpha}.$

\item  $m_\alpha$ is the number of base functions appearing in model $\alpha$.

\item  $B_{j\alpha}$ is the amplitude of base function $G_{j\alpha}$.

\end{itemize} 

The measurements are described by:

$$
d_l=M_\alpha (t_l)+e_l 
$$

where $e_l$ is the noise realized in measurement $l$. Given the model and a
parameter set, the difference between a datum value and the value (of this
datum) predicted by the model is assigned to the noise.

We discuss here the case of Gaussian assignment to the noise, and discuss 
the arising computational implications. Indeed, the assignment of Gaussian probability
 distribution to the noise is a special case, but 
a very important one: arising from maximum entropy considerations, 
it is the proper encoding of our ignorance regarding non-systematic effects
with known/finite variance {\cite {Jayens12}}.

The probability of getting the value of a datum point, given the Gaussian
probability distribution of the noise, its variance $\sigma _s,$ and all the
model parameters, is: 
$$
p(d_1/\overrightarrow{B}_\alpha\overrightarrow{\tau }_\alpha
 M_\alpha \sigma _sI)=\frac 1{%
\sqrt{2\pi \sigma _s^2}}\exp (-\frac 1{2\sigma
_s^2}(d_1-\sum_{j=1}^mB_{j\alpha}G_{j\alpha}
(t_1,\overrightarrow{\tau _{j\alpha}}))^2) 
$$

For the sake of clarity, from now on we omit the index $\alpha$ that signified the model $M_\alpha $
and the explicit symbol of the model $M_\alpha $. We do everything inside the
model under consideration.

We deal with the general case of multiple measurements. That is, each
time-series was sampled more than once.
The noise is assumed uncorrelated, and thus the probability of getting this
data is the product of the probability of getting each of the data points: 
$$
p(D/\overrightarrow{\tau }\overrightarrow{B}\sigma
_sI)=\prod_i^n\prod_l^{n_l}p(d_{il}/\overrightarrow{\tau }\overrightarrow{B}%
\sigma _sI) 
$$

$$
=(2\pi \sigma _s^2)^{-\frac{n\cdot n_l}2}\exp (-\frac 1{2\sigma
_s^2}\sum_{i=1}^n\sum_{l=1}^{n_l}(d_{il}-\sum_{j=1}^mB_jG_j(t_i,
\overrightarrow{\tau _j})^2)= 
$$

\begin{equation}
=(2\pi \sigma _s^2)^{-\frac{n\cdot n_l}2}\exp (-\frac Q{2\sigma _s^2}) 
\label{baseEq}
\end{equation}

Where

$$
Q=nn_l\overline{d^2}-2\sum_{j=1}^m\sum_{i=1}^nB_jG_j(t_i,
\overrightarrow{\tau _j})\cdot
\sum_{l=1}^{n_l}d_{il}+n_l\sum_{j=1}^m\sum_{k=1}^mg_{jk}B_jB_k 
$$

$$
g_{jk}=\sum_{i=1}^nG_j(t_i,\overrightarrow{\tau _j})\cdot 
G_k(t_i,\overrightarrow{\tau _j}) 
$$

$$
\overline{d^2}=\frac 1{n\cdot n_l}\sum_{i=}^n\sum_{l=1}^{n_l}d_{il}^2 
$$

and

\begin{itemize}
\item  $m$ -- Number of model base functions.

\item  $n$ -- Number of sampling points.

\item  $n_l$ -- Number of measurements of each time-series
\end{itemize}

Equation (\ref{baseEq}) is the key expression to be computed in any
calculation of probabilities in time series problems with Gaussian
noise.

In order to understand the implications of having multiple measurements and
the relationship between the part of noise arising from the misfits of the 
data to the model and the part of noise arising from the non-systematic effects, we let: 
$$
\sigma ^2=\frac{\sigma _s^2}{n_l} 
$$

and use this quantity from now on. So,

\begin{equation}
\label{modpLikeNonOrhto}
p(D/\overrightarrow{\tau }\overrightarrow{B}\sigma
I)=(2\pi n_l\sigma ^2)^{-\frac{n\cdot n_l}2}\exp (-\frac Q{2\sigma ^2}) 
\end{equation}

$$
Q=n\overline{d^2}-2\sum_{j=1}^m\sum_{i=1}^nB_jG_j(t_i,\overrightarrow{\tau _j})
\overline{d_i}%
+\sum_{j=1}^m\sum_{k=1}^mg_{jk}B_jB_k 
$$

Where:

$$
\overline{d_i}=\frac 1{n_l}\sum_{l=1}^{n_l}d_{il} 
$$

We see that $\overline{d_i}$, the average of the data points for every
sampling time $i$, is  sufficient for finding the maximum of the
likelihood. It is not sufficient to establish the absolute value of the
likelihood at any point of the parameter space.

Nevertheless, in case $\sigma $ is known, $\overline{d_i}$ is sufficient
for any calculation of the parameters' values, since the
likelihood dependence on the more detailed distribution of the data is
through a constant factor,$\overline{d^2}$.

As common sense suggests, if $\sigma $ is unknown, the extent to which the
data points are distributed will determine the probability distribution of $%
\sigma $. The probability distribution of $\sigma $, in turn, will not
influence the values of the parameters at maximum likelihood, but will
determine the width of the likelihood and of the probability
distributions for the parameters.

In order to compute the probability of the model $P(M/DI)$ for the model 
comparison
and estimation of the parameters, one must integrate the likelihood given in
Equation(\ref{modpLikeNonOrhto}) over all the model parameters.

Having the base functions orthogonal, in the sense that:

$$
g_{jk}=\sum_{i=1}^n G_j(t_i)\cdot G_k(t_i)=\delta _{jk} 
$$

will enable us to analytically integrate the likelihood over the linear
parameters. Moreover, in case the base functions are orthonormal, the
expectation values of their amplitudes are given directly by the projection
of the data on the base functions \cite{BretBook}.

In spite of the general dependence of the base functions on nonlinear
parameters, the base functions will be, in many important problems, almost
orthogonal. For example, in the problem of evenly sampled multiple
stationary frequencies, with cosines and sines as the base functions, the
off diagonal terms of $g_{jk}$ will be negligible as long as the frequencies
are well separated \cite{BretBook}.

In this paper we demonstrate the application of Bretthorsts' formulation
for a rather pathological problem, in which the matrix $g_{jk}$ is never
orthogonal. The computational difficulties, arising from the existence 
of nonlinear parameters in our problem, lead
us to consider computation-reduction tools: 

First, arranging the models into
a multi-dimensional model space, in which a search can be performed 
instead of impractical calculation of the probability of all the
models. Second, the Bretthorst formulation teach us how to 
analytically integrate over the linear parameters, but we still have
to integrate over the nonlinear parameters numerically or through 
approximations. Numerical integrations are usually impractical in spaces of many 
nonlinear parameters. We are left with the task of finding the 
transformation of the likelihood (as a function of the nonlinear
parameters) into forms that can be approximated with satisfactory
precision.

Analyzing a multiple-measurements problem, we also note about the ``noise"
and its meaning in the Bayesian framework. In cases of
multiple-measurements, we can rarely estimate the noise by using
 statistics like the Standard Deviation.

\section{Posing the Problem}

\subsection{The Data}

Using a 32 echo CPMG MR imaging sequence on a 1.5T GE clinical
MR scanner, spin-spin relaxation $(T_2)$ decay curves were acquired from the
brains of 11 normal humans \cite{AleKenVivo} . A decay curve is the collection 
of the amplitudes of the same pixel in the image through 32 consecutive images.
Accordingly, each decay curve contains amplitudes of 32 points in time, from
10 to 320ms (Figure (\ref{intdata})). The partition to specific tissues was
assigned by a neurologist on the images of ten brains. For each of twelve
tissues, five of white matter and seven of gray matter, 800-5000 decay
curves had been collected. The eleventh brain's data provides the new data
for the classification stage.

\begin{figure}[t] 
\setlength{\unitlength}{1.00mm} 
\begin{picture}(110,65) 
\includegraphics{intdat.eps} 
\end{picture} 
\caption{Sample of the Raw Data. Different symbols signify data taken from different
humans.} 
\label{intdata} 
\end{figure}

It is clear from the Bayesian standpoint, that we lose information by using
this data instead of the raw signal collected by the scanner. The available
data is the result of the imaging reconstruction, including FFT, and thus
suffers from FFT artifacts. Moreover, after assigning the specific tissues,
the localization of the pixels' data is lost, preventing us from taking
possible tissue assignment errors and inter-tissue contamination into
account. Nevertheless, in our formulation of the problem, we regard this
data as the raw data. The information loss will contribute to the ``noise''.

\subsection{Background Information}

\subsubsection{The Set of Models}
\label{ModelSet}

The mechanism that produced the decay curves is modeled by: 
$$
d_i=\int_0^\infty f(T_2)\cdot e^{-\frac{t_i}{T_2}}dT_2+[a+[bt+[ct^2]]]+d%
\cdot (-1)^i\cdot e^{-\frac{t_i}{70}}+Noise 
$$
Where i=1..32

\begin{itemize}
\item  The $T_2$ Distribution $f(T_2)$ is expected to be composed of a few
discrete components. These components are parameterized by their amplitude,
decay rate, and possibly width: 
$$
f(T_2)=\sum_m^{j-k}a_m\delta (T_2m-T_2)\ +\ \sum_n^ka_n\frac 1{\sqrt{2\pi
(width)^2}}\cdot e^{-\frac{(T_{2n}-T_2)^2}{2(width)^2}} 
$$

A possible $T_2$ distribution is given in Figure (\ref{sampdist}).

\begin{figure}%[t]
\setlength{\unitlength}{1.00mm}
\begin{picture}(127,50)
\includegraphics{sampdist.eps}
\end{picture}
\caption{Example of possible $f(T_2)$, the distribution of the $T_2$ decay rate.}
\label{sampdist}
\end{figure}

\item  Polynomial components may exist, and their amplitudes are not known.

\item  The third term, an Alternating-Echo-Refocusing (AER) exists, its
amplitude not known.
\end{itemize}

So, our models can be arranged in a 3-D model space. Model $M_{jkl}$ will
have

\begin{itemize}
\item  {\bf j} Decaying components. We assume no more than 7 such components.

\item  {\bf k} Decaying components with wide distribution of the decay rate.

\item  {\bf l} Polynomial components. We assume no more than 3 such
components.
\end{itemize}

\subsubsection{Noise}

The noise, in the Bayesian interpretation, is not a `random' process, but
merely a process where our ignorance regarding the producing mechanism is
such that its effect on the data can not be anticipated exactly. Any known
behaviour of the producing mechanism can, in principle, be extracted from the
noise and incorporated into the model. The noise will include any effects,
systematic or not, that are present in the data but not in the model. The
non-systematic effects in our case, are expected to be of finite variance
but are otherwise not known. Accordingly, we assign the noise a Gaussian
distribution with unknown variance\cite{jaynes45}.

\subsubsection{Prior Information}

We assume no preference to any model in the model space, and ignorance
regarding the values of the models' parameters. The numerical representation
of this ignorance is assigned by considering the amplitudes of any of our
base functions as location parameters (leading to flat priors), and the
decay rates and widths of components as scale parameters (leading to
Jeffreys' priors). One may argue that the ignorance regarding some of the
parameters should be represented differently; nonetheless, as long as we 
keep our priors uninformative, the influence of our ignorance 
representation on the final results is negligible. 

The priors for all the models are assigned equal, by
indifference.

\subsection{The Questions}

The questions we ask are always regarding the posterior probability (or {\it 
{pdf}}) of the proposition we are interested in.

\subsubsection{Model Selection and Parameter Estimation}

For each tissue, which model $M_{jkl}$ has the highest posterior probability 
$p(M_{jkl}/D_{tissue}I)$? Given that model, what are the the most probable
values of the models' parameters?

\subsubsection{Classification}

After inferring about the $T_2$ distribution of each of the tissues, using a
collection of data sets, we are interested in the following question: Given
a new set of data $d_{new}$, we want to find the tissue it came from. We are
looking for the tissue $i$ that will maximize the probability $%
p(C_i/d_{new}D_iI)$ where $C_i\equiv $ ``The new data was produced by the
same mechanism that produced the old data set $D_i$.'' This mechanism is
described by a model (a functional form) and values of the model
parameters.

\subsection{The Process: Model Selection and Parameter Estimation}

Section (\ref{ModelSet}) defined the set of models to be tested.
However, we do not wish to calculate the needed posterior probability $%
p(M_{jkl} /DI) $ for all these models. Though their number is finite, the
computation time needed for computing all of them will render our task
impractical. Our algorithm computation time is polynomial in the number of linear
parameters and exponential in the number of nonlinear parameters
\footnote{This is since the number of linear parameters determines the dimension 
of matrices to be diagonalized, and the number of nonlinear parameters, on the other hand, 
determines the dimension of space to be searched.}. We
need to find the most probable model without calculating all the models, and
in particular we want to avoid unnecessary computations of models with many
nonlinear parameters. In order to achieve that, we describe our model space as 
3-D discrete space, using the j,k,l
coordinates defined in Section (\ref{ModelSet}), and search this 
space for the most probable model.
 
A flow chart of the search path is given in Figure (\ref{modflEps}), and the model
space with example of search in it is given in Figure (\ref{modspace}).

\begin{figure}%[t]
\setlength{\unitlength}{1.00mm}
\begin{picture}(127,60)
\includegraphics{flowchrt.eps}
\end{picture}
\caption{Search Path Flow Chart. The algorithm
is looking for the maximum probability in the space of models.}
\label{modflEps}
\end{figure}

\begin{figure}[h] 
\setlength{\unitlength}{1.00mm} 
\begin{picture}(127,90) 
\includegraphics{modspace.eps} 
\end{picture} 
\caption{Search in the Space of Models. 
The second-best model is not necessarily a neighbour of the best model.} 
\label{modspace} \end{figure}

We now turn to the more detailed computation of each model.
The posterior probability of model $M_{jkl}$ is given by:

\begin{equation}
p(M_{jkl}/D_iI)=\int \int p(M_{jkl}\Theta _{jkl}\Phi _{jkl}/d_iI)d\Theta
_{jkl}d\Phi _{jkl} 
\label{modsel}
\end{equation}

$$
\alpha \int \int p(d_i/M_{jkl}\Theta _{jkl}\Phi _{jkl}I)d\Theta _{jkl}
d\Phi_{jkl}
$$

Where $\Theta _{jkl}$ and $\Phi _{jkl}$ are the linear and nonlinear 
parameter sets of model $M_{jkl}$, respectively. 

We use the formulation of Bretthorst\cite{BretBook}: The
integration over the linear parameters is done via an orhtogonalizing
transformation of the base functions. This transformation depends on the
nonlinear parameters.

In order to integrate over the non-linear parameters by quadratic
approximation, we transform the non-linear parameters to a new set of
parameters in which the Log[likelihood] is sufficiently quadratic around its
peak. In our case, this transformation is the logarithm (Figure (\ref{trans})).

\begin{figure}[b] 
\setlength{\unitlength}{1.00mm} 
\begin{picture}(127,80) 
\includegraphics{trans.eps} 
\end{picture} 
\caption{Transformation of the Log[Likelihood] into Quadratic Form.} 
\label{trans} 
\end{figure}

In Figure (\ref{modpar}) we bring a sample of the results of applying
Equation (\ref{modsel}) to the data collected from different tissues. The 
figure shows
the models that had been selected for particular tissues and the estimated
parameters for these tissues. The search for the models always started from
a simple model of one decaying exponent. The credible regions for the
parameters cannot be seen in the figure. 

\begin{figure}[b] 
\setlength{\unitlength}{1.00mm} 
\begin{picture}(110,100) 
\includegraphics{estires.eps} 
\end{picture} 
\caption{Results of Model Selection and Parameter Estimation for White Matter
Tissues (left) and Gray Matter Tissues (right).} 
\label{modpar} 
\end{figure}

\subsection{The Process: Classification}

The posterior probability of $C_i$ is just the summation, over all possible
models (and the values of their parameters) , 
of the likelihood of the model (and the values of its parameters) in 
light of the new data, weighted by
the probability of that model (and the values of its parameters) 
given the old data set $D_{tissue}$: 
$$
p(C_i/d_{new}D_{tissue}I) 
$$
$$
\alpha \sum_{jkl}\int \int p(M_{jkl}\Theta _{jkl}\Phi
_{jkl}/D_{tissue}I)\cdot p(d_{new}/\Theta _{jkl}\Phi _{jkl}M_{jkl}I)\
d\Theta _{jkl}\ d\Phi _{jkl} 
\label{proclass}
$$

We use Equation (\ref{proclass}) to classify each pixel in a 32-image-set,
generating a synthetic image, in which different classified tissues 
are represented by different grey levels (Figure (\ref{class})).

\begin{figure}[b] 
\setlength{\unitlength}{1.00mm} 
\begin{picture}(127,50) 
\includegraphics{mri.ps} 
\end{picture} 
\begin{picture}(127,50) 
\includegraphics{class.ps} 
\end{picture} 
\caption{Results of Classification: The MRI Image (left) and the Classified Synthetic Image (right). Each gray level in the Synthetic Image corresponds to certain tissue. In the data used for this work, the variance among humans was larger than the differ
ence between tissues. The resulting image fails to classify closely behaving tissues, though the discrimination between White and Gray Matter is reasonable.}
\label{class} 
\end{figure} 

\subsection{Discussion}

\subsubsection{Search in a multi-dimensional model space}

In order to efficiently find the most probable model, avoiding unnecessary
computation of complex models, we formulated a search routine in a 3-D model 
space, including
stopping conditions. The arrangement of the models into the particular 
was not forced by the nature of the problem and is a matter of decision.  
The search routine used in this work found to be robust
for the $T_2$ problem as formulated, but is definitely an {\it {ad-hoc}}
development. The general problem of establishing a search routine in the
model space is difficult. It is not clear what transformation should be used
in order to simplify the topology and avoid many local maxima, as
demonstrated in Figure (\ref{modspace}). This transformation will in general
depend on the priors of the models, the priors for the parameters values,
the structure of the data, and the initial model attributes that span the
space.

\subsubsection{Integration over nonlinear parameters}

In general, the nonlinear parameters cannot be eliminated by analytical
integration, and for cases of many coupled parameters, 
cannot be integrated numerically either. In most cases, we have to approximate
the likelihood by another function, for which analytical integration is feasible.

The normal form of the likelihood which results from the Gaussian 
assignment to the noise leads us to look for quadratic 
approximations: The behaviour of Log[Likelihood] as function of any 
linear parameter is quadratic. In fact, the Bretthorst method for
orthogonalization and integration over the linear parameters is equivalent 
to the integration over quadratic approximation of the function.  

For the nonlinear parameters, the problem is reduced to the general problem of
linear approximation;
that is, finding the transformation of the nonlinear
parameters that will bring the Log[likelihood] to a form which is
sufficiently quadratic around its peak.
The transformation used in the problem of the $T_2$ distribution was not found
in any consistent procedure, but was suggested by the nature of the specific 
problem as analyzed in previous, `frequentist', works. The possible 
resulting errors in the final posterior probabilities for the models were 
calculated and a ``safety valve routine" was applied in order to ensure detection
and correction of cases in which the Log[likelihood] did not transform 
into a quadratic enough function.

\subsubsection{Noise}

Our likelihood is a product of terms of the form: 
$$
p(d_i/\sigma \Theta \Phi I)=\frac 1{\sqrt{2\pi \sigma_s ^2}}\cdot e^{-\frac{%
(M_i-d_i)^2}{2\sigma_s ^2}} 
$$

In many case we are tempted to approximate $\sigma_s$ using the standard 
deviation of the data. We recognize that in the frame-work of probability 
theory as logic the `noise' characterize the
deviation of the individual data points from model and not from the 
mean of the data as the STD does. Nevertheless, assuming our model is close enough 
to the data mean, we hope it will serve as a sufficient approximation. 

Alas, {\it{In the case of multiple measurements, $\sigma_s $ does not characterize the
deviation of the individual data points from our model (``The Noise
Level'')}}. In the case of systematic effects not accounted for by the model,
the estimation of $\sigma_s $ will diverge in the limit of many measurements.
In this limit, the quantity $\frac {\sigma_s} {\sqrt{n}}$ (assigned the 
symbol $\sigma$ in this work) will converge to a 
finite number, characterizing the magnitude of the systematic effects in 
the data not accounted for by the model.


\begin{thebibliography}{9}

\bibitem{BretBook}  G.~L. Bretthorst. \newblock {\em Bayesian Spectrum
Analysis and Parameter Estimation}. \newblock Lecture Notes in
Statistics:48. Springer-Verlag, 1988.

\bibitem{JaynesBook}  E.T.Jaynes. \newblock {\em Probability theory - the
Logic of Science}. \newblock 1993.

\bibitem{Myron}  M. Tribus. \newblock {\em Rational Descriptions, Decisions
and Design}. \newblock Pergamon Press, Oxford, 1969.

\bibitem{Jayens12}  E.~T. Jaynes. \newblock {\em Bayesian Spectrum and Chirp
Analysis}. \newblock Maximum Entropy and Bayesian Spectral Analysis and
Estimation Problems. D. Reidel ed., Dordrecht-Holland, 1987.

\bibitem{AleKenVisual}  A. MacKay, K.Whittall, J. Adler, D. Li, D. Paty, D.
Graeb\newblock {\em In Vivo visualization of myelin water in brain by
magnetic resonance}. \newblock Mag. Res. in Med., 31:673--677, 1994.

\bibitem{AleKenVivo}  A.~L.~MacKay, K.~P. Whittall, R. A. Nugent, D. Graeb,
I. Lees, D.K.B. Li\newblock {\em In vivo }$T_2${\em \ relaxation in normal
human brain}. \newblock Mag. Res. in Med., Aug. 1994.

\bibitem{AleKenQuant}  Kenneth~P. Whittall and Alexander~L. MacKay. 
\newblock {\em Quantitative interpretation of NMR relaxation data.} 
\newblock J. of Mag. Res., 84:134--152, 1989.

\bibitem{jaynes45}  E.T.Jaynes. \newblock {\em Prior probabilities}. 
\newblock IEEE Transactions on Systems science and Cybernetics,
SSC-4:227--241, 1968.

\bibitem{BretII}  G.~L. Bretthorst. \newblock {\em Bayesian analysis II:
Signal detection and model selection.} \newblock J. of Mag. Res,
88:552--570, 1990.


\end{thebibliography}
\end{document}